\documentclass[]{jfm}

\usepackage[english]{babel}

\usepackage{graphicx}
\usepackage{natbib}
\usepackage{hyperref}
\usepackage[utf8]{inputenc}
\hypersetup{
    colorlinks = true,
    urlcolor   = blue,
    citecolor  = black,
}

\newcommand{\RomanNumeralCaps}[1]
\linenumbers

\usepackage{amsmath,latexsym,amssymb,amsfonts}
\usepackage{csquotes}
\usepackage{bm}

\usepackage[dvipsnames]{xcolor}

\newcommand{\vect}[1]{\boldsymbol{#1}}

\newcommand{\R}{\vect{r}}

\newcommand{\Intd}{\mathrm{d}}

\definecolor{AirForceBlue}{rgb}{0., 0.34, 0.66}

\title{Diffusiophoretic propulsion of an isotropic active colloidal particle near a finite-sized disk embedded in a planar fluid-fluid interface} 

\author{Abdallah Daddi-Moussa-Ider\aff{1}, Andrej Vilfan\aff{1,2}, \and Ramin Golestanian\aff{1,3}} 

\affiliation{
  \aff{1}Max Planck Institute for Dynamics and Self-Organization (MPIDS), 37077 G\"ottingen, Germany
  \aff{2}Jo\v{z}ef Stefan Institute, 1000 Ljubljana, Slovenia
  \aff{3}Rudolf Peierls Centre for Theoretical Physics, University of Oxford, Oxford OX1 3PU, United Kingdom
  }

\begin{document}
\maketitle

\begin{abstract}

	Breaking spatial symmetry is an essential requirement for phoretic active particles to swim  at low Reynolds number. 
	This fundamental prerequisite for swimming at the micro-scale is fulfilled either by chemical patterning of the surface of active particles or alternatively by exploiting geometrical asymmetries to induce chemical gradients and achieve self-propulsion.
	In the present manuscript, a far-field analytical model is employed to quantify the leading-order contribution to the induced phoretic velocity of a chemically homogeneous isotropic active colloid near a finite-sized disk of circular shape resting on an interface separating two immiscible viscous incompressible Newtonian fluids.
	To this aim, the solution of the phoretic problem is formulated as a mixed-boundary-value problem which is subsequently transformed into a system of dual integral equations on the inner and outer domains.
	Depending on the ratio of different involved viscosities and solute solubilities, the sign of phoretic mobility and chemical activity, as well as the ratio of particle-interface distance to the radius of the disk, the isotropic active particle is found to be either repelled from the interface, attracted to it, or reach a stable hovering state and remains immobile near the interface.
	Our results may prove useful in controlling and guiding the motion of self-propelled phoretic active particles near aqueous interfaces.

\end{abstract}

\begin{keywords}
	Biological fluid dynamics, low-Reynolds-number flows, active matter, swimming/flying	
\end{keywords}

\section{Introduction}
\label{sec:intro}

The emerging field of active soft matter physics has recently gained considerable attention in the biophysics and bioengineering communities~\citep{lauga09,Golestanian2011, elgeti15, bechinger16, zottl16, illien17, RG-LesHouches,gompper20}.
Over the past few years, there has been a mounting research interest in designing and developing self-propelling microswimmers as they are set forth as model systems for understanding the fundamentals of out-of-equilibrium phenomena in physiology and cellular biology.
Synthetic man-made self-propelled active swimmers are capable of propelling themselves autonomously through a liquid by converting the energy extracted from their surrounding host environment into useful mechanical work.
They are thought to hold great promise for future biomedical and clinical applications such as drug delivery, biopsy, precision nanosurgery, diagnostic histopathology, and transport of curative substances to tumour cells and inflammation sites~\citep{gao14, wang13}. 
Suspensions of active components have been shown to lead to the emergence of a wealth of intriguing collective phenomena and fascinating spatiotemporal patterns.
Prime examples include the motility-induced phase separation~\citep{tailleur08, buttinoni13, speck14,soto+golestanian14}, propagating density waves and swarms~\citep{gregoire04, mishra10, menzel12,Cohen:2014,saha-prx:2020}, and the emergence of active meso-scale turbulence~\citep{uchida2010synchronization,wensink12pnas, dunkel13, heidenreich14, thampi2014vorticity, kaiser14, doostmohammadi17, doostmohammadi2018active,berta-prx:2021}.

Phoretic self-propulsion is a well-established mechanism of choice in active matter research \cite{RG-LesHouches}.
Unlike most of the remotely actuated swimmers that fully rely on an external field to propel themselves through aqueous media~\citep{dreyfus2005microscopic, ghosh2009controlled, tierno2008controlled, wang14acoustic}, self-phoretic swimmers stand apart since they can achieve intrinsic self-propulsion solely by exploiting local physico-chemical interactions with the surrounding fluid medium, while inherently fulfilling the force- and torque-free constraints required for swimming at the micron scale~\citep{Golestanian:2005,Golestanian:2007}. 
Phoretic active colloids can be set to motion through an effective slip velocity resulting from a local concentration gradients induced via surface chemical reactions~\citep{Derjaguin1947,RG-LesHouches,sharifi13, michelin14, ibrahim2017multiple, moran2017phoretic,julicher2009generic, brady2011particle}.
Various theoretical works have been devoted to uncovering the effect of particle shape~\citep{Golestanian:2007,popescu2010phoretic, nourhani2016geometrical, michelin2017geometric, ibrahim2018shape, katsamba2020slender, poehnl2020axisymmetric} and geometric confinement~\citep{popescu2009confinement, uspal16, mozaffari16, bayati2019dynamics} on the behaviour and dynamics of self-phoretic particles. The collective behaviour of multiple phoretic particles has been studied in a number of different contexts \citep{Golestanian:2012,saha+golestanian14,Kranz-trail-PRL:2016,Gelimson-trail-PRL:2016,Saha2019,varma2019modeling}.

Breaking the spatial symmetry is a main prerequisite to achieve phoretic self-propulsion at the low Reynolds numbers~\citep{Golestanian:2005,Golestanian:2007}.
From an experimental standpoint, the most commonly followed approach to fulfil this physical requirement consists of chemically patterning the surface of active colloidal particles \citep{howse07, Ebbens2010,Ebbens2012,Ebbens2014,das15, Kurzthaler:2018, campbell2019experimental}.
An alternative route to accomplishing self-phoretic locomotion without the need for micro-patterning is based on exploiting geometrical asymmetries to induce chemical gradients~\citep{michelin2015autophoretic}.
Indeed, isotropic self-phoretic particles can swim by means of phoretic and hydrodynamic interactions with other inert (non-motile) particles by forming dynamical clusters of anisotropic geometry \cite{SotoGolestanianPRL:2014,SotoGolestanianPRE:2015,varma2018clustering,Agudo-Canalejo2019, nasouri2020prl, nasouri2020jfm}. Meanwhile, \cite{lisicki2016phoretic} demonstrated that internal phoretic flows can be induced solely by geometric asymmetries of chemically homogeneous surfaces.

In the present contribution, we employ a far-field approach to examine the diffusiophoretic motion of an isotropic active colloidal particle of spherical shape positioned in the vicinity of a finite-sized disk resting on a planar interface separating two immiscible fluid media. We formulate the phoretic problem as a classical mixed-boundary-value problem, which we subsequently transform into a system of dual integral equations on the inner and outer domain boundaries. We perform an explicit calculation of the hydrodynamic flow field by making use of the Lorentz reciprocal theorem in fluid mechanics to yield an analytical expression of the induced phoretic velocity normal to the interface. More importantly, we find that the active particle can be repelled from or attracted to the interface depending on the particle-interface distance relative to the disk size, the ratios of fluid viscosities and solubilities of species in the two media bounded by the interface in addition to the sign of the phoretic mobility and chemical activity. Consequently, the self-phoretic swimming behaviour can be controlled by adequately tuning the physical and geometrical properties of the system.

\section{Problem formulation}

\begin{figure}
	\centering
	\includegraphics[scale=1.55]{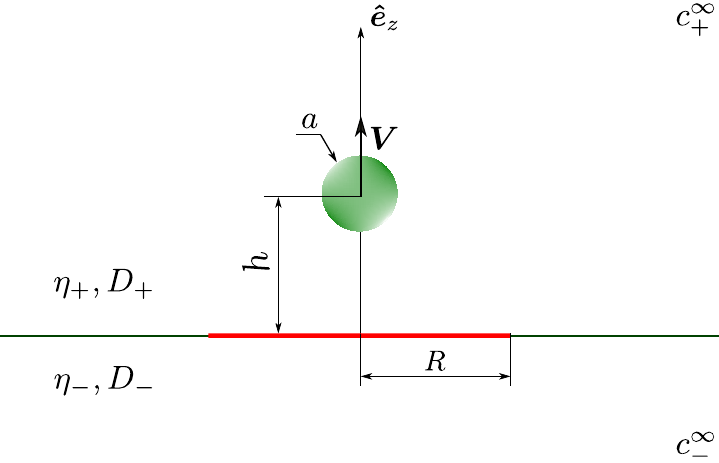}
	\caption{(Colour online) Schematic illustration of the system
          setup.  An active isotropic particle of radius~$a$ located
          at position~$h$ on the axis of an impermeable no-slip
          disk of radius~$R$.  The
          disk is embedded in an interface between two mutually
          immiscible fluids with dynamic viscosities
          $\eta_\pm$.  We denote by $D_\pm$ the diffusion
          coefficient of the chemical and by~$c_\pm^\infty$ the
          equilibrium far-field concentration of the solute in each fluid
          domain.  }
	\label{GraphicalIllustration}
\end{figure}

We examine the axisymmetric motion of a spherical active colloidal particle near a thin impermeable circular disk, resting on a flat fluid-fluid interface. The interface extends infinitely in the plane $z = 0$.
The active particle is coated with a catalyst that promotes a chemical reaction converting fuel molecules to products.
We denote by the subscripts~$(+)$ the parameters and variables in the upper fluid domain above the interface, for which $z > 0$, and by the subscript~$(-)$ the parameters and variables in the region occupied by the fluid underneath the interface, for which $z < 0$.
We assume that the fluids on both domains are Newtonian and incompressible with uniform dynamic viscosities~$\eta_\pm$. 
An infinitely thin disk of radius~$R$ is positioned within the plane $z=0$ separating the two immiscible fluids.
The active particle of radius~$a$ is fully immersed in the upper fluid medium at position~$h$ on the symmetry axis of the disk.
We denotes by~$D_\pm$ the diffusion coefficients of the fuel molecules in each fluid compartment. 
In the following, we employ a far-field approach to describe the induced hydrodynamic and concentration fields. We note that the effect of thermal noise as well as number fluctuations in the chemical field have been ignored throughout our calculations \cite{Golestanian:2009}.

\subsection{Equations for the concentration field}

We suppose that the surface of the active particle emits or absorbs the solute with a uniform flux density $Q$ such that,
\begin{equation}
\left. -D_+ \, \frac{\partial c_+}{\partial r} \right|_{r = a} = Q \, , \label{BC:Q}
\end{equation}
wherein $Q$ can be positive or negative depending on whether the catalytic reaction is associated with a production (emission) or annihilation (absorption) of the solute.

At low P\'eclet numbers, the advection of the solute by the flow is negligible in relation to diffusion.
Under these conditions, the evaluation of the solute distribution can be decoupled from that of the fluid flow.
Accordingly, the stationary concentrations in the upper and lower domains are described by the Laplace equation,
\begin{equation}
	\boldsymbol{\nabla}^2 c_\pm(\R) = 0 \, . \label{LaplaceEq}
\end{equation}
Equation~\eqref{LaplaceEq} is subject to the boundary conditions of fixed concentration $c_\pm^\infty$ far away from the active particle as $\left| \R \right|\to \infty$. The surface of the finite-sized disk imposes a no-flux boundary condition
\begin{equation}
\left. \frac{\partial c_\pm}{\partial z} \right|_{z=0} = 0 \qquad\qquad (\rho < R) \, . \label{BC:no-flux}
\end{equation}
Outside the disk, the fluid-fluid interface requires a continuous chemical flux,
\begin{equation}
\left. D_+ \, \frac{\partial c_+}{\partial z} = D_- \, \frac{\partial c_-}{\partial z} \right|_{z=0} \qquad\qquad (\rho > R) \, .
\label{BC:FluidInterfaceGrad}
\end{equation}
We define the dimensionless number
\begin{equation}
\lambda = \frac{D_-}{D_+} = \frac{\eta_+}{\eta_-} \, ,
\end{equation}
assuming that the Stokes-Einstein relation is valid for diffusion in both domains.

In addition, we allow a discontinuous concentration field at the fluid-fluid interface as a result of different solubilities of the chemical in the two fluid media via~\citep{dominguez16}
\begin{equation}
\left. \ell c_+ = c_- \right|_{z = 0} \qquad\qquad (\rho > R) \, ,  \label{BC:FluidInterfaceConcentration}
\end{equation}
where $\ell = c_-^\infty / c_+^\infty$ is a dimensionless ratio that determines the equilibrium concentrations in the unperturbed fluid.

\subsection{Phoretic propulsion}

We consider the frequently employed assumption of a short-range potential between the particle and solute molecules such that mutual interactions are limited to a thin boundary layer the active particle~\citep{Golestanian:2005, Golestanian:2007}.
Accordingly, the slip velocity at the surface of the active colloid, $\mathcal{S}_\mathrm{P}$, can be obtained from the tangential gradient of the concentration field as
\begin{equation}
	\vect{v}_\mathrm{S} = \left. -\mu \boldsymbol{\nabla}_\parallel c_+ \right|_{\mathcal{S}_\mathrm{P}} \, , 
	\label{SlipVelocity}
\end{equation}
with~$\mu$ denoting the phoretic mobility that is defined from the profile of the local interaction potential between the particle and solute molecules. 
In addition, $\boldsymbol{\nabla}_\parallel (\cdot) = r^{-1} \partial (\cdot) / \partial \theta \, \vect{e}_\theta $ stands for the tangential gradient along the surface of the sphere.

\section{Solution for the concentration field}

In the far-field limit, the active particle can conveniently be approximated as a point source. We express the solution of the Laplace equations for the concentration field in both fluid domains as a sum of a direct contribution $C$ and the contributions of the boundary or the flux across the boundary  $c_\pm^*$:
\begin{equation} 
	c_+ = c_+^\infty + C + c_+^* \, , \qquad\qquad	c_- = c_-^\infty + c_-^* \, , \label{ImageSolutionInit}
\end{equation}
Here $C$ is the solution of Eq.~\eqref{LaplaceEq} in an unbounded fluid medium subject to the constant flux boundary condition at the surface of the active particle stated by Eq.~\eqref{BC:Q}.
Specifically,
\begin{equation}
C (\rho, z) = K \left( \rho^2 + \left(z-h\right)^2 \right)^{-\frac{1}{2}} \, , \label{Eq:C}
\end{equation}
where we have defined the length scale $K = a^2 Q/D_+$.

Here $c_\pm^*$ are the complementary (also often referred to as the image) solutions that are required to satisfy the boundary conditions prescribed at the fluid-fluid interface as well as at the surface of the finite-sized disk. Being harmonic functions, we express the image solutions in terms of Fourier-Bessel integrals of the form
\begin{equation}
c_\pm^* (\rho, z) = \int_0^\infty A_\pm (q) J_0(q \rho) e^{- q |z|} \, \Intd q \, , \label{ImageSolutionForm}
\end{equation}
with $J_0$ denoting the zeroth-order Bessel function of the first
kind. The wavenumber-dependent functions $A_\pm (q)$ will be
subsequently determined from the underlying boundary conditions.  

\subsection{Formulation of the dual integral equations}
The equations for the inner problem ($\rho<R$) can readily be obtained by inserting Eq.~\eqref{ImageSolutionInit} into Eq.~\eqref{BC:no-flux} prescribing the no-flux boundary condition at the surface of the finite-sized disk, to obtain 
\begin{subequations} \label{InnerProblem}
	\begin{eqnarray}
	\int_0^\infty q A_+(q) J_0 (q \rho) \, \Intd q 
	&=& \left. \frac{\partial C}{\partial z} \right|_{z=0} \qquad\qquad (\rho < R) \, , \label{InnerProblem1} \\
	\int_0^\infty q A_-(q) J_0(q \rho) \, \Intd q 
	&=& 0 \hspace{2.7cm} (\rho < R) \, . \label{InnerProblem2}
	\end{eqnarray}
\end{subequations}

On the other hand, the equations for the outer problem ($\rho>R$) follow from applying the boundary conditions imposed at the fluid-fluid interface given by Eqs.~\eqref{BC:FluidInterfaceGrad} and \eqref{BC:FluidInterfaceConcentration},
\begin{subequations} \label{OuterProblem}
	\begin{eqnarray}
	\int_0^\infty q \left( A_+(q) + \lambda A_-(q) \right) J_0 (q \rho) \, \Intd q 
	&=& \left. \frac{\partial C}{\partial z} \right|_{z=0} \qquad\qquad (\rho > R) \, , \label{OuterProblem1} \\
	\int_0^\infty \left( A_-(q) - \ell A_+(q) \right) J_0(q \rho) \, \Intd q 
	&=& \left. \ell C \right|_{z=0} \hspace{1.69cm} (\rho > R) \, . \label{OuterProblem2}
	\end{eqnarray}
\end{subequations}

Equations~\eqref{InnerProblem} and~\eqref{OuterProblem} form a system of dual integral equations for~$A_\pm (q)$ on the inner and outer domain boundaries.
Analytical solution of such type of integral equations with Bessel function kernels can often be obtained by employing the theory of Mellin transforms~\citep{tranter51, titchmarsh48book}. 
However, we choose to follow an alternative strategy using the well-established solution approach described by \cite{sneddon60} and \cite{copson61}.
In particular, we will show that the present system of dual integral equations can eventually be reduced to classical Abel integral equations, amenable to inversion in explicit form.
We note that this solution approach has previously frequently been utilized to solve diverse flow problems involving finite-sized boundaries.
These include the determination of the viscous flow field induced by various types of singularities acting near an elastic disk possessing shear and bending deformation modes~\citep{daddi19jpsj, daddi2020asymmetric}, near a no-slip disk~\citep{kim83, daddi2020dynamics, daddi2021steady}, or between two coaxially positioned rigid disks of the same size~\citep{daddi2020axisymmetric}.
The present approach has also been employed to determine the electrostatic potential in a circular plate capacitor with disks of different radii~\citep{paffuti2016circular}.

\subsection{Solution of the dual integral equations}

By combining the equations for the inner problem given by Eq.~\eqref{InnerProblem} and invoking Eq.~\eqref{OuterProblem1}, it follows that
\begin{equation}
	\int_0^\infty q \left( A_+(q) + \lambda A_-(q) \right) J_0 (q \rho) \, \Intd q 
	= \left. \frac{\partial C}{\partial z} \right|_{z=0} 
\end{equation}
applies for all values of~$\rho$.
Accordingly, the Hankel transform can be applied on both sides of the equation to obtain
\begin{equation} 
	A_+(q) + \lambda A_-(q) = \int_0^\infty \left. \frac{\partial C}{\partial z} \right|_{z=0} \, J_0(q \rho) \rho \, \Intd \rho \, ,
\end{equation}
which leads us upon inserting the expression of $C(\rho, z)$ given by Eq.~\eqref{Eq:C} to
\begin{equation}
	A_+(q) + \lambda A_-(q) = K e^{-q h} \, . \label{SolutionFormFromHankel}
\end{equation}

To satisfy the equations for the outer domain for which, we choose a solution of the integral form
\begin{equation} 
	A_-(q) - \ell A_+(q)   = \ell K e^{-q h} + \int_0^R f(t) \sin (q t) \, \Intd t \qquad (\rho > R) \, , \label{SolutionForm}
\end{equation}
which clearly satisfy Eqs.~\eqref{OuterProblem2} because $\mathcal{L}_0^0(\rho, t) = 0$ for $t<R<\rho$ (see Appendix~\ref{appendix:Lmn}, Eq.~\eqref{K10_K20}).
Solving Eqs.~\eqref{SolutionFormFromHankel} and \eqref{SolutionForm} for $A_\pm (q)$ yields 
\begin{subequations} \label{SolutionFormFinal}
	\begin{eqnarray}
	A_+ (q) &=& K \Lambda_1  e^{-q h} + \lambda M (q) \, , \\
	A_- (q) &=& K \Lambda_2 e^{-q h} - M (q) \, ,
	\end{eqnarray}
\end{subequations}
where we have defined $\Lambda_1 = \left( 1-\lambda\ell \right) / \left( 1+\lambda\ell \right)$ and~$\Lambda_2 = 2 \ell / \left( 1+\lambda\ell \right)$.
Moreover,
\begin{equation}
	M (q) = -\frac{1}{1+\lambda\ell } \int_0^R  f(t) \sin (q t) \, \Intd t   \, . \label{Mplus}
\end{equation}
Substituting the expressions of~$A_\pm(q)$ given by Eqs.~\eqref{SolutionFormFinal} into either integral equations for the inner problem given by Eq.~\eqref{InnerProblem} yields 
\begin{eqnarray}
	\int_0^\infty q M (q) J_0 (q \rho) \, \Intd q &=& \frac{\Lambda_2 Kh}{\left( \rho^2 + h^2 \right)^\frac{3}{2}} \qquad\qquad (\rho < R) \, . \label{Inner1}
\end{eqnarray}

To proceed further, we employ integration by parts to obtain
\begin{eqnarray}
\int_0^R q f(t) \sin (q t) \, \Intd t = \varphi(q)  +\int_0^R f'(t) \cos (q t) \, \Intd t \, ,
\end{eqnarray}
wherein $\varphi(q) = f(0) - f(R) \cos (q R)$.
Correspondingly, Eq.~\eqref{Inner1} can be expressed 
\begin{equation}
\int_0^R \mathcal{L}_1^0(\rho, t) f_2'(t) \, \Intd t = G(\rho) \qquad\qquad (\rho < R) \, , \label{Eq:f1_f2Prime}
\end{equation}
where we have interchanged the order of integration with respect to $t$ and~$q$ and defined for convenience
\begin{equation}
G(\rho) = -\frac{2\ell K h}{\left( \rho^2 + h^2 \right)^\frac{3}{2}}  - f(0) \mathcal{L}_1^0(\rho, 0) + f(R) \mathcal{L}_1^0(\rho, R) \, .
\end{equation}

It follows from Eq.~\eqref{K10_K20} that $\mathcal{L}_1^0(\rho, 0) = 1/\rho$ and~$\mathcal{L}(\rho, R) = 0$ (because $\rho < R$ holds in the inner domain).
Then, Eq.~\eqref{Eq:f1_f2Prime} simplifies to
\begin{equation}
\int_0^\rho \frac{f'(t) \, \Intd t}{\left( \rho^2 - t^2 \right)^\frac{1}{2}} = -\frac{2\ell Kh}{\left( \rho^2 + h^2 \right)^\frac{3}{2}} - \frac{f(0)}{\rho} \, . \label{Abel1}
\end{equation}

Since $f(0)$ is required to vanish for Eq.~\eqref{Abel1} to be defined at $\rho = 0$, the resulting equation for $f(t)$ reduces to a classical Abel integral equation.
The latter represents a special form of the Volterra equation of the first kind possessing a weakly singular kernel~\citep{carleman21, smithies58, anderssen80}.
It admits a unique solution if and only if the radial function on the right-hand side is a continuously differentiable function~\citep{carleman22, tamarkin30, whittaker96}.
Its solution is obtained as
\begin{equation}
	f(t) = -\frac{4}{\pi} \frac{\ell K t}{t^2 + h^2} \, . \label{f2}
\end{equation}

By inserting the latter expression of~$f(t)$ into Eq.~\eqref{Mplus}, we obtain
\begin{equation}
	M (q) = \Lambda_2 K \int_0^R \frac{2}{\pi} \frac{t \sin (q t)}{t^2+h^2} \, \Intd t \, .
\label{MPlus_Final}
\end{equation}

In particular, in the limit $R \to \infty$ corresponding to an infinitely extended impermeable wall, we get $M (q) = \Lambda_2 K e^{-q h}$, leading to $A_+(q) = K e^{-q h}$ and $A_-(q) = 0$.

Finally, by inserting the expression for $M (q)$ into Eqs.~\eqref{SolutionFormFinal} and substituting the resulting expressions of $A_\pm(q)$ into Eq.~\eqref{ImageSolutionForm}, the solutions for the concentration field are obtained as
\begin{subequations} \label{ImageSolutionConcentrationFinal}
  \begin{eqnarray}
    c_+^* (\rho, z) &=& \Lambda_1 C(\rho, -z) + \lambda \, \tilde{c} (\rho, z) \, , \\
    c_-^* (\rho, z) &=& \Lambda_2 C(\rho, +z) -  \tilde{c} (\rho, z) \, .
  \end{eqnarray}
\end{subequations}
The first term in each expression is a simple image $C(\rho,\pm z)$ that describes the effect of the fluid-fluid interface. The second term $\tilde c(\rho,z)$ can be interpreted as the field induced by an effective source dipole distribution in the boundary that compensates the flux across the interface, such that the superposition of both terms obeys the zero-flux boundary condition. 
Here, the contribution resulting from the presence of the impermeable disk can be written in the form of a definite integral as
\begin{equation}
	\tilde{c} (\rho, z) = \Lambda_2 K \int_0^R \frac{2}{\pi} \frac{t \, \mathcal{U} (\rho, z, t)}{t^2 + h^2} \, \Intd t \, ,
\end{equation}
where we have defined 
\begin{equation}
\mathcal{U} (\rho, z, t) = \int_0^\infty J_0 (q \rho) \sin (q t) e^{-q |z|} \, \Intd q \, .
\end{equation}

\begin{figure}
	\centering
	\includegraphics[scale=0.8]{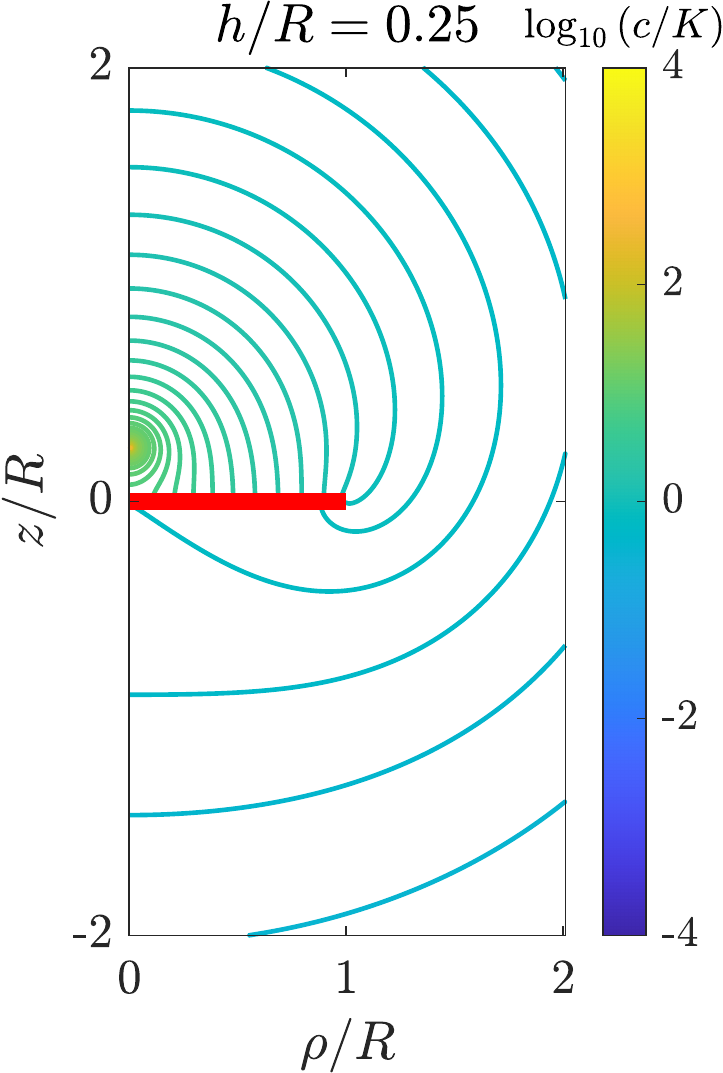}~~~~
	\includegraphics[scale=0.8]{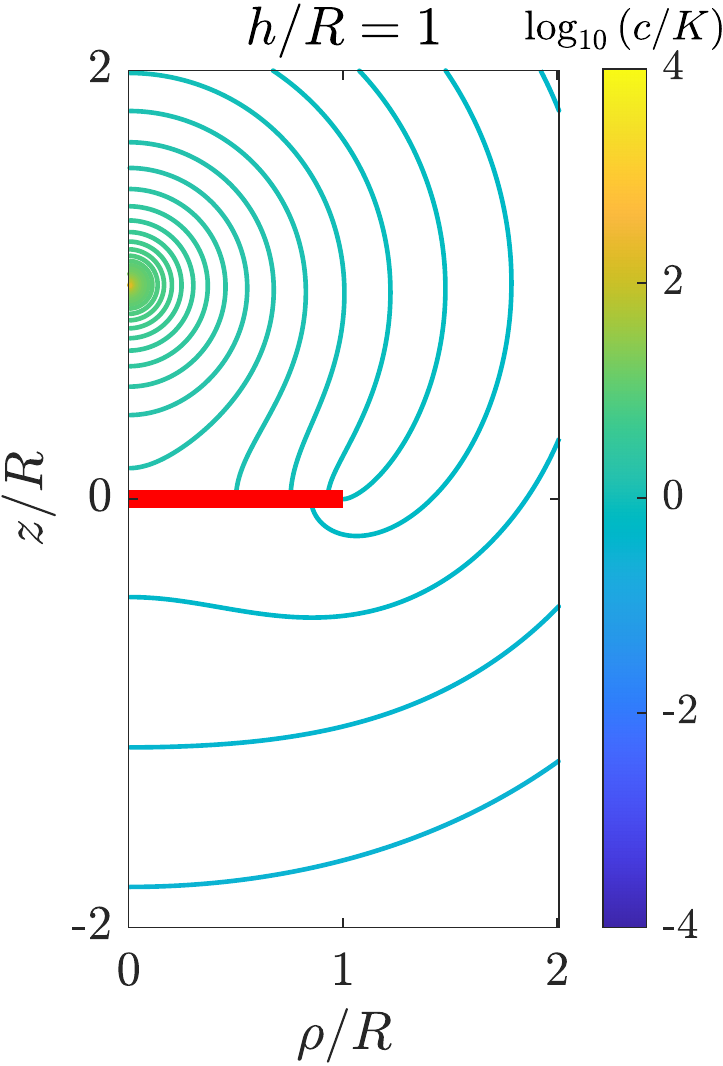}
	\put(-355,237){{\Large (a)}}
	\put(-175,237){{\Large (b)}}
	\caption{(Colour online) 
		Contour plots of the scaled concentration field around an active particle positioned at (a) $h/R = 0.25$ and (b) $h/R = 1$ on the axis of a finite-sized impermeable disk of radius~$R$ (shown in red) resting on a fluid-fluid interface with viscosity ratio $\lambda = 1$ and solubility ratio $\ell = 1$.}
	\label{contour1}
\end{figure}

\begin{figure}
	\centering
	\includegraphics[scale=0.8]{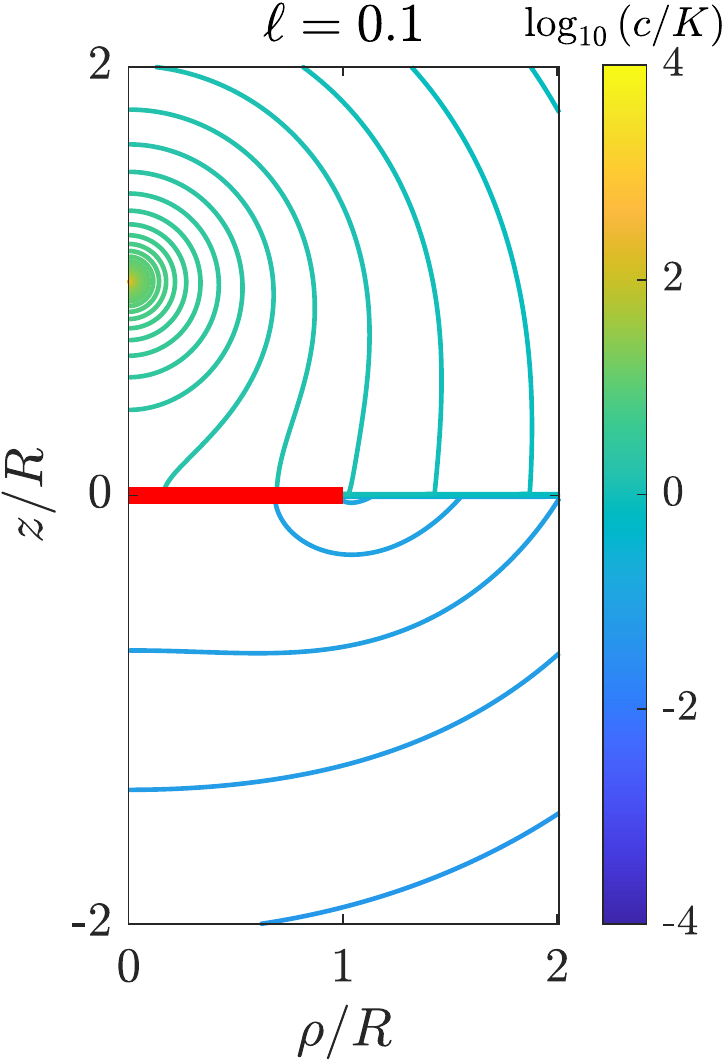}~~~~
	\includegraphics[scale=0.8]{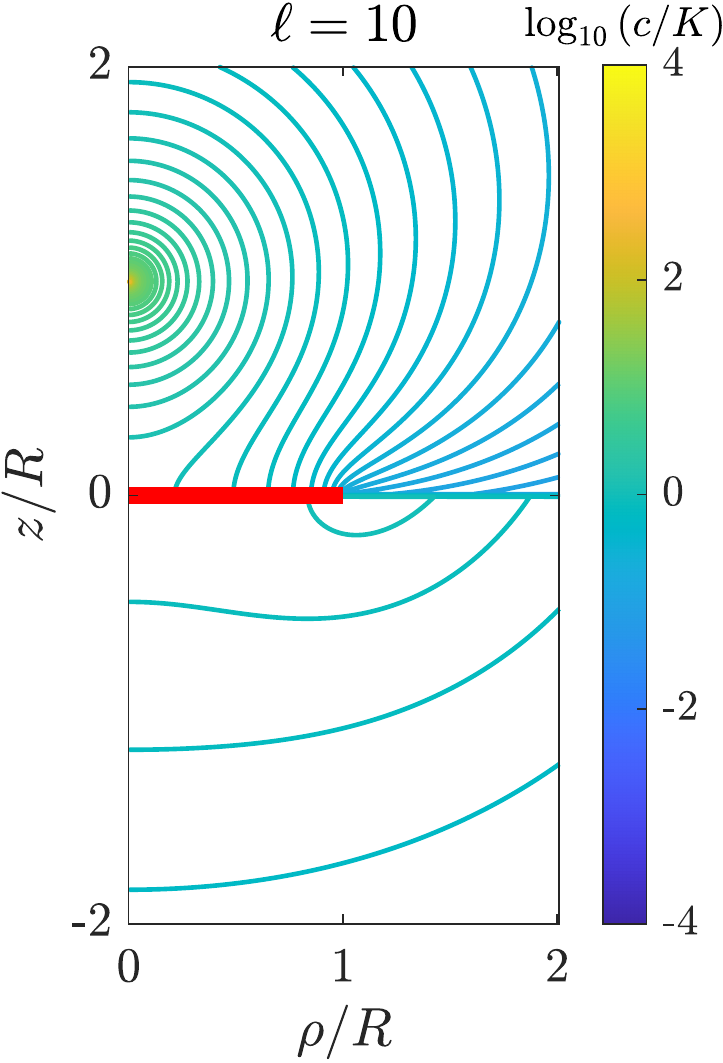}
	\put(-355,237){{\Large (a)}}
	\put(-175,237){{\Large (b)}}
	\caption{(Colour online) 
		Contour plots of the scaled concentration field around an active particle positioned at $h/R = 0.5$ for $\lambda = 1$ and solubility ratios (a) $\ell = 0.1$ and (b) $\ell = 10$.}
	\label{contour2}
\end{figure}

We will show in the sequel that an analytical evaluation of the latter improper integral is possible by invoking concepts from complex analysis.
Using the substitution $u = q \rho$, we obtain
\begin{equation}
\mathcal{U} (\rho, z, t) = \rho^{-1} \operatorname{Im} \left\{ \int_0^\infty J_0 (u) e^{-su} \, \Intd u \right\} 
\end{equation}
with $s = \left( |z| - it \right)/\rho$. 
By recalling the Laplace transform of $J_0 (u)$ which is given by $\left( 1+s^2 \right)^{-\frac{1}{2}}$, we obtain
\begin{equation}
\mathcal{U} (\rho, z, t) = \operatorname{Im} \left\{ \left( \rho^2 + \left( |z| - it \right)^2 \right)^{-\frac{1}{2}} \right\} .
\end{equation}
The latter can further be expressed in the form by evaluating the imaginary part as
\begin{equation}
\mathcal{U} (\rho, z, t) = \left( \left( U-V \right)/2 \right)^\frac{1}{2} / U \, , 
\end{equation}
where we have defined
\begin{equation}
U = \left( \left(  \rho^2 + z^2 + t^2 \right)^2 - \left(2\rho t\right)^2 \right)^\frac{1}{2} \, , \qquad
V =  \rho^2 + z^2 - t^2 \, .
\end{equation}

In the special case $R\to\infty$, representing an infinitely extended impermeable wall, the image solution  $c_+^*(\rho,z)=C(\rho,-z)$, $c_-^*(\rho,z)=0$ is recovered by noting that (see Appendix \ref{sec:appendix-integral} for the proof)
\begin{equation}
	\int_0^\infty \frac{t \, \mathcal{U} (\rho, z, t)}{t^2+h^2} \, \Intd t = \frac{\pi}{2} \left( \rho^2 + \left( |z|+h \right)^2 \right)^{-\frac{1}{2}} \, . \label{EqtoBeProven}
\end{equation}

Exemplary contour plots illustrating the lines of equal concentration, also sometimes called isopleths, are shown in Fig.~\ref{contour1} for two different singularity positions above the interface while keeping the viscosity and solubility ratios equal to one. 
The presence of the disk introduces an asymmetry in the form assumed by the lines of equal concentration owing to the no-flux boundary condition imposed at the surface of the disk.
Analogous contour plots are shown in Fig.~\ref{contour2} upon varying of the ratio of solubility between the two media.

\section{Phoretic velocity}

Al low Reynolds numbers, the dynamics of the viscous Newtonian fluids in the two fluid domains is governed by the steady Stokes equations~\citep{kim13}
\begin{subequations} \label{StokesGleischung}
	\begin{eqnarray}
	\boldsymbol{\nabla} \cdot \vect{v}_\pm &=& 0 \, , \\
	-\boldsymbol{\nabla} p_\pm + \eta_\pm \boldsymbol{\nabla}^2 \vect{v}_\pm &=& 0 \, ,   
	\end{eqnarray}
\end{subequations}
where $\vect{v}_\pm$ denote the flow velocity and $p_\pm$ the pressure.

\subsection{Lorentz reciprocal theorem}

In lieu of directly solving the governing equations for fluid motion for the prescribed boundary conditions, we follow an alternative route based on the Lorentz reciprocal theorem \citep{lorentz96amsterdam, kuiken1996general, happel12, masoud2019reciprocal, stone96}.
This approach has extensively been used in the context of phoretic swimming to determine the propulsion velocity of chemically active colloids suspended in an unbounded fluid medium~\citep{Golestanian:2005,Golestanian:2007,popescu16, oshanin2017active}, close to a planar no-slip wall~\citep{crowdy2013wall, uspal15, ibrahim15, yariv16, papavassiliou2015many}, near a chemically patterned surface~\citep{uspal2018shape, popescu2017chemically, uspal2019active}, or to compute the stresslet field induced by active swimmers~\citep{lauga16prl}.
Further, the reciprocal theorem has been adapted to describe the phoretic interaction of two active Janus particles~\citep{bayati2016dynamics, sharifi16, nasouri2020prl, nasouri2020jfm}, or to investigate the behaviour of a self-propelled active particle in a complex fluid~\citep{lauga2014locomotion, elfring2017force, datt2017active}.

According to the reciprocal theorem, two distinct solutions of the Stokes equations $\left( \vect{v}, \boldsymbol{\sigma}  \right)$ and $\left( \hat{\vect{v}}, \hat{\boldsymbol{\sigma}} \right)$ within the same fluid domain~$\mathcal{D}$ bounded by a surface~$\mathcal{S}$ are related to each other via
\begin{equation}
\int_\mathcal{S} \vect{n} \cdot \boldsymbol{\sigma} \cdot \hat{\vect{v}}  \, \Intd S = \int_\mathcal{S} \vect{n} \cdot \hat{\boldsymbol{\sigma}} \cdot \vect{v} \, \Intd S  \, , \label{LRT}
\end{equation}
with~$\vect{n}$ denoting the unit vector normal to the surface~$\mathcal{S}$ pointing outward.
In the following, unhatted and hatted quantities will be used to refer to the flow properties in the main and model (also called sometimes auxiliary or dual) problems, respectively.

By decomposing the fluid domain on both sides of the fluid-fluid interface into upper and lower domains, it can be shown that the natural continuity of the tangential components of the fluid velocity and traction at the plane $z = 0$ implies a vanishing contribution to the surface integral over the fluid-fluid interface~\citep{sellier2011migration}.
Moreover, since the fluid velocity is assumed to vanish at infinitely as well as at the surface of the no-slip disk, surface integration reduces to that over the surface of the colloid~\citep{malgaretti2018self}.
Consequentially, the reciprocal theorem takes precisely a form analogous to that expressed for a colloidal particle in an unbounded fluid medium.

As a model problem, we consider the axisymmetric motion of a chemically inert (passive) spherical particle dragged through the upper fluid with velocity~$\hat{\vect{V}}$ by a steady externally applied force $\hat{\vect{F}} = \hat{F} \, \vect{e}_z$.
Correspondingly, $\left. \hat{\vect{v}} \right|_{\mathcal{S}_\mathrm{P}} = \hat{\vect{V}}$ is constant over the surface of the particle and can thus be taken out of the surface integral.
Further, because the active particle is force free, the resulting integral on the left-hand side of Eq.~\eqref{LRT} identically vanishes.
At the surface of the active colloid, the fluid velocity can be decomposed as $\left. \vect{v} \right|_{\mathcal{S}_\mathrm{P}} = \vect{V} + \vect{v}_\mathrm{S}$, with~$\vect{V} = V \, \vect{e}_z$ standing for the net drift velocity of the active particle and the slip velocity~$\vect{v}_\mathrm{S}$ is given by Eq.~\eqref{SlipVelocity}.
Therefore, the translational phoretic velocity of the chemically active particle can be obtained from
\begin{equation}
\vect{V} \cdot \hat{\vect{F}} = \mu \oint_{\mathcal{S}_\mathrm{P}} \hat{\vect{f}} \cdot \boldsymbol{\nabla}_\parallel c_+ \, \Intd S \, ,
\end{equation}
with~$\hat{\vect{f}} = \vect{n} \cdot \hat{\boldsymbol{\sigma}}$ denoting the traction at the surface of a sphere in the model problem.
Up to $\mathcal{O} \left( \epsilon^2 \right)$ it is given by $\hat{\vect{f}} = \left( 4\pi a^2 \right)^{-1} \hat{\vect{F}}$.
By employing the transformations $\rho = r\sin\theta$ and $z = h + r\cos\theta$ and noting that $\vect{e}_z \cdot \vect{e}_\theta = -\sin\theta$, the induced phoretic velocity can eventually be expressed up to $\mathcal{O} \left( \epsilon^3 \right)$ in terms of an integral over the polar angle as
\begin{equation}
V = -\frac{\mu}{a} \int_{-1}^1 \zeta \, c_+ (r=a, \zeta) \, \Intd \zeta \, , 
\label{InducedPhoreticVelocity}
\end{equation}
where we have used integration by parts and introduced the change of variable $\zeta = \cos\theta$.
Correspondingly, the induced phoretic velocity is given by the first moment of concentration~\citep{michelin2013spontaneous}.

\subsection{Leading-order contribution to the phoretic velocity}

At this point, we have derived the solution of the diffusion equation for a point-source singularity acting on the symmetry axis of a finite-sized disk resting on a fluid-fluid interface.
We will next make use of this solution to determine the induced phoretic velocity of an active colloidal particle with isotropic surface activity.
To calculate the leading order contribution, we restrict ourselves to the point-particle approximation which is valid when $a \ll h$.

The image solutions derived above and given by
Eqs.~\eqref{ImageSolutionConcentrationFinal} satisfy exactly the
boundary conditions prescribed at the surface of the disk and at the
fluid-fluid interface.  However, they disturb the constant flux
condition imposed at the surface of the active particle.  To overcome
this shortcoming, a series of images needs to be incorporated so as to
satisfy the boundary conditions in an alternative manner up to a
desired accuracy.   The next-order contribution to the concentration field
consists of an axisymmetric source dipole
\begin{equation}
  \label{eq:sourcedipole}
  c_+^\mathrm{SD} = A \, \frac{\partial}{\partial z} \left( \rho^2 + \left(z-h\right)^2 \right)^{-\frac{1}{2}} \, , 
\end{equation}
with
\begin{equation}
	A = -\frac {a^3}{2} \lim\limits_{(\rho, z) \to (0, h)} \frac {\partial c_+^*}{\partial z} \, .
\end{equation}
In this way, the boundary conditions are satisfied up to $\mathcal{O} \left( \epsilon^3 \right)$ both at the particle surface and the interface.
Specifically,
\begin{subequations}
	\begin{eqnarray}
	c_+ &=& c_+^\infty + C + \Lambda_1 \bar{C} + \lambda \tilde{c} + a \,\frac{\partial C}{\partial z} \frac{\epsilon^2}{8} \left(\Lambda_1 + \lambda \Lambda_2 \Gamma \right)  , \\
	c_- &=& c_-^\infty + \Lambda_2  C - \tilde{c} +  a \, \frac{\partial C}{\partial z} \frac{\epsilon^2}{8} \Lambda_2 \left( 1 + \Gamma  \right) ,
	\end{eqnarray}
\end{subequations}
where we have defined the abbreviation~$\bar{C} (\rho, z) = C(\rho, -z)$. 
Furthermore, $\Gamma$ is obtained so as to fulfil the constant flux boundary condition imposed at the surface of the active particle to $\mathcal{O} \left( \epsilon^3 \right)$.
It is explicitly given by
\begin{equation}
	\Gamma = \frac{2}{\pi} \left( \frac{Rh \left( R^2-h^2 \right)}{\left( R^2+h^2 \right)^2} + \arctan \left( \frac{R}{h} \right) \right) .
\end{equation}

Then, by making use of Eq.~\eqref{InducedPhoreticVelocity} providing the induced phoretic velocity, we obtain
\begin{equation}
	V = -\frac{\mu Q}{D_+} \frac{\epsilon^2}{1+\lambda\ell} \left( \frac{1-\lambda\ell}{4} + \lambda\ell J(\xi) \right)
+ \mathcal{O} \left( \epsilon^3 \right)	, \label{PhoreticVelocity}
\end{equation}
wherein~$\xi = h/R$ and
\begin{equation}
J(\xi) = \frac{1}{\pi} \left( \frac{\xi \left( 1-\xi^2 \right)}{\left( 1+\xi^2 \right)^2 }+ \arctan \left( \xi^{-1} \right)  \right)
\end{equation}
is a monotonically decreasing function of~$\xi$ varying between $1/2$ and~0.
In the limit $\xi \ll 1$, we obtain
\begin{equation}
V = -\frac{\mu Q}{4D_+} \, \epsilon^2 \left( 1 - \frac{16}{3\pi} \, \lambda \Lambda_2 \xi^3  \right) + \mathcal{O} \left( \xi^5 \right) .
\end{equation}
In particular, we recover the leading-order far-field contribution to the induced phoretic velocity in the limit of an infinitely extended no-slip wall as originally obtained by \cite{ibrahim15} and later generalized by \cite{yariv16} for both remote and near-contact configurations using a first-order kinetic model of solute absorption.

\begin{figure}
	\centering
\includegraphics[scale=1]{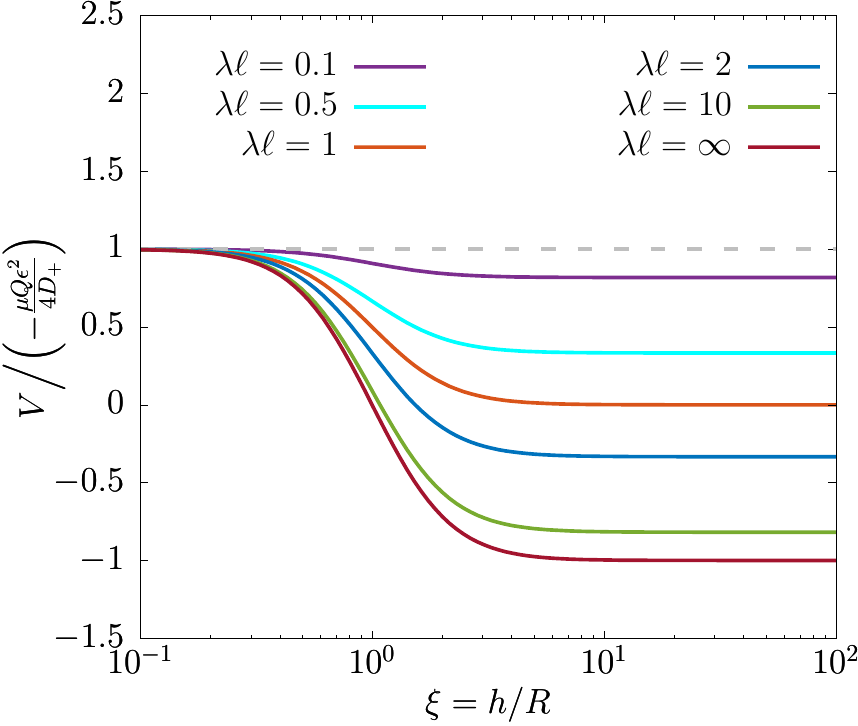}
	\caption{(Colour online) Variation of the scaled induced phoretic velocity near a finite-sized disk resting on a fluid-fluid interface as given by Eq.~\eqref{PhoreticVelocity} versus the dimensionless number $\xi = h/R$ for various values of~$\lambda \ell$.
	Horizontal dashed line corresponds to the situation of an infinite wall.}
	\label{SwimmingSpeed}
\end{figure}

Figure~\ref{SwimmingSpeed} shows in a semilogarithmic scale the evolution of the scaled phoretic velocity as given by Eq.~\eqref{PhoreticVelocity} as a function of $\xi = h/R$.
Results are presented for six different values of~$\lambda\ell$ which span the most likely values for fluid-fluid interfaces to be expected for a wide range of practical situations. 
The scaled velocity amounts to its maximum value as $\xi \to 0$ and decreases monotonically with~$\xi$ to reach the value corresponding to a fluid-fluid interface given by $\left( 1-\lambda\ell \right) / \left( 1+\lambda\ell \right)$ in the limit $\xi \to \infty$.
The active particle is found to be repelled from the interface or attracted to it, depending on the sign of the phoretic mobility~$\mu$, the flux~$Q$, as well as the values of the dimensionless parameters~$\lambda\ell$ and~$\xi$.
Under some circumstances, the particle remains in a stationary hovering state in which it acts as a micropump.

The phoretic speed keeps the same sign over the whole range of values of~$\xi$ when $\lambda\ell \le 1$.
Correspondingly, the particle is repelled from the interface for $\mu Q < 0$ and attracted for $\mu Q > 0$.
This behaviour is analogous to what has been earlier reported for diffusiophoresis near an infinite no-slip wall~\citep{ibrahim15, yariv16}.
In contrast to that, the induced speed can eventually vanishes and changes sign when $\lambda\ell > 1$.
By equating Eq.~\eqref{PhoreticVelocity} to zero and solving for~$\xi$, we find that the phoretic velocity vanishes at a unique value $\xi = \xi_0$ given by the solution of
\begin{equation}
	\frac{1}{\xi_0} = \tan \left( \frac{\pi}{4} \left( 1 - \frac{1}{\lambda\ell} \right) - \frac{\xi_0 \left( 1-\xi_0^2 \right)}{\left( 1+\xi_0^2 \right)^2} \right) \qquad\qquad (\lambda\ell > 1) \, . \label{xi0}
\end{equation}
Accordingly, for $\lambda\ell > 1$, the active particle is repelled from the interface if $\mu Q \left( \xi-\xi_0 \right) > 0$ and attracted to it if $\mu Q \left( \xi-\xi_0 \right) < 0$.

For $\lambda\ell \gg 1$, we obtain the scaling relation
\begin{equation}
	\xi_0 = 1 + \frac{\pi}{4} \left(\lambda\ell\right)^{-1} + \mathcal{O} \left( \left(\lambda\ell\right)^{-2} \right) \, .
\label{xi0Slope}
\end{equation}

In particular, it follows from Eq.~\eqref{InducedPhoreticVelocity} that the induced phoretic velocity near a fluid-fluid interface is obtained as
\begin{equation}
V = -\frac{\mu Q}{4 D_+} \frac{1-\lambda\ell}{1+\lambda\ell} \, \epsilon^2 + \mathcal{O} \left( \epsilon^3 \right) .
\label{PhoreticVelocityFluidInterface}
\end{equation}

\begin{figure}
	\centering
\includegraphics[scale=1]{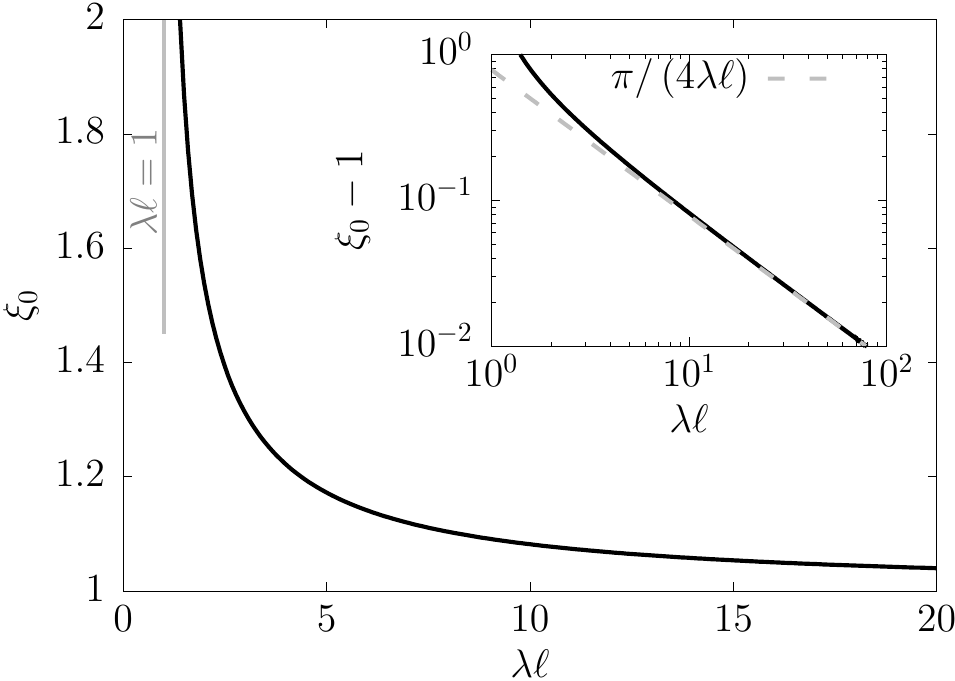}
	\caption{Variation of $\xi_0$ defined by Eq.~\eqref{xi0} corresponding to a vanishing induced phoretic velocity versus~$\lambda\ell$.
	The inset shows the scaling behaviour around~$\lambda\ell \to \infty$ as given by Eq.~\eqref{xi0Slope}.}
	\label{XiNot}
\end{figure}

In Fig.~\ref{XiNot} we present the variation of~$\xi_0$ by numerically solving Eq.~\eqref{xi0} using standard computational techniques.
We remark that $\xi_0$ asymptotically approaches infinity as $\lambda\ell \to 1$ and monotonically decreases before reaching a minimum value of one as $\lambda\ell \to \infty$.
The linear scaling behaviour predicted by Eq.~\eqref{xi0Slope} is shown in using a log-log scale in the inset.

Up to now, we have obtained the leading-order contribution to the phoretic velocity of an active isotropic colloid suspended in the vicinity of a circular disk of finite size settling on a surface separating two fluids.
A power series solution for the induced phoretic velocity can in principal be obtained perturbatively by considering additional singular fields in the concentration field.
However, due to their complexity and the intricate form of the image solution, accounting for additional singularities is rather delicate and laborious.

\begin{figure}
	\centering
\includegraphics[scale=1]{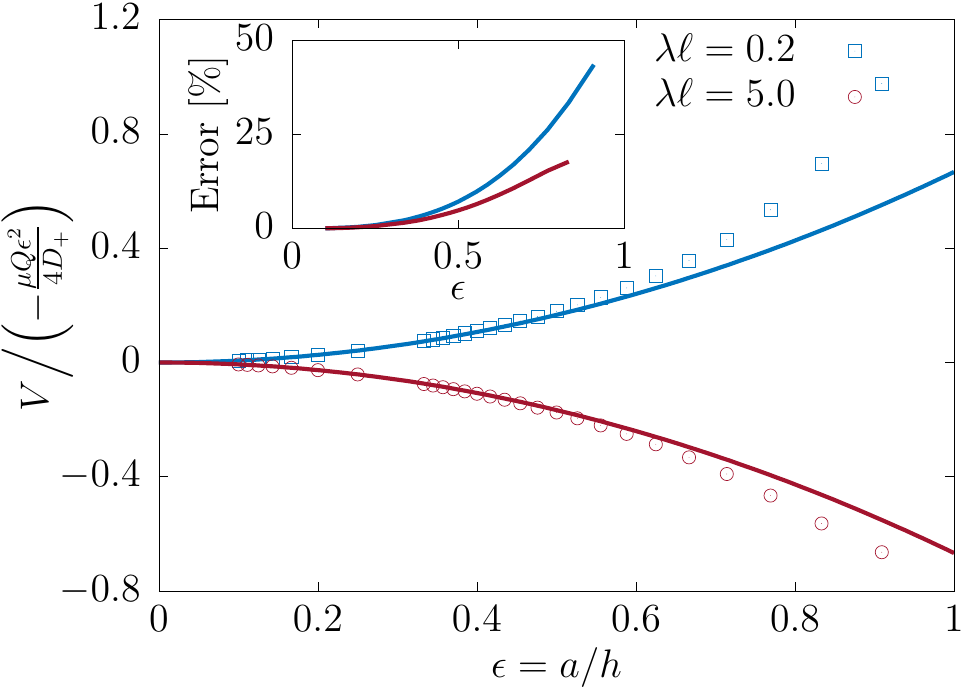}
	\caption{(Colour online) Scaled induced phoretic velocity near an infinitely extended fluid-fluid interface (in the absence of the disk) as a function of the dimensionless particle size $\epsilon = a/h$ for two different values of~$\lambda\ell$.
	Symbols represent the exact results obtained using bipolar coordinates \citep{malgaretti2018self} and solid lines indicate the far-field solution derived in the present work given by Eq.~\eqref{PhoreticVelocityFluidInterface}.
	Here, the viscosity ratio between the two media is $\lambda = 1$.
	The inset shows the relative percentage error, which is of the order $\propto \epsilon^3$. 
	\label{PaoloFig}
	}
\end{figure}

Figure~\ref{PaoloFig} shows a comparison of the scaled phoretic velocity near a fluid-fluid interface as obtained by means of bipolar coordinates (symbols) recently reported by~\cite{malgaretti2018self} and the far-field expression given by Eq.~\eqref{PhoreticVelocityFluidInterface}.
Results are plotted versus the dimensionless ratio~$\epsilon$ of particle radius to distance from the interface for $\lambda\ell = 0.2$ (blue) and $\lambda\ell = 5$ (red) while the viscosity ratio is kept $\lambda = 1$.
Good agreement is obtained between the exact analytical solution and the simplistic far-field expression derived in the present work.
In particular, both approaches capture the same underlying physical behaviour on whether the particle moves toward the interface or away from it.
As shown in the inset, the far-field approach leads to a relative percentage error smaller than~$10\%$ when $\epsilon < 0.5$.
The error monotonically increases as the particle gets closer to the interface, reaching for $\epsilon = 0.9$ about $20\%$ for $\lambda\ell = 5$ and $40\%$ for $\lambda\ell = 0.2$.

\section{Conclusions}

To summarize, we have presented a far-field analytical theory addressing the axisymmetric autophoretic motion of an isotropic active particle suspended in a viscous Newtonian fluid medium near a rigid disk embedded in a planar fluid-fluid interface.
We have formulated the solution for the concentration field induced by a point-source singularity as a standard mixed-boundary-value problem which we have then reduced into a classical Abel integral equation amenable to analytical inversion.
By making use of Lorentz reciprocal theorem, we have obtained an analytical expression for the leading-order far-field contribution to the induced phoretic velocity, thereby elevating the need to solve for the hydrodynamic flow field explicitly.

On the one hand, we have shown that for $\lambda\ell < 1$, the induced velocity normal to the interface depends solely on the phoretic mobility and chemical activity and is found to be independent of the system geometrical properties.
Specifically, the case $\mu Q < 0$ corresponds to repulsion from the interface while $\mu Q > 0$ corresponds to attraction. 
On the other hand, we have shown that for $\lambda\ell > 1$, there exists a size ratio~$\xi_0$ for which the active particle reaches a steady motionless hovering state above the interface.
Beyond this state, the active particle is found to be repelled from (attracted to) the interface depending on whether $\mu Q \left( \xi - \xi_0 \right)$ is positive (negative).

The present analytical developments are based on a far-field description of the phoretic and hydrodynamic fields.
They rely on the assumption that the active particle is small relative to its distance from the interface.
As a perspective, it would be of interest to assess the appropriateness and accuracy of the point-particle approximation employed in this work by direct comparison with fully resolved numerical boundary integral solution.
In addition, it would be worthwhile to extend our analytical approach to address the behaviour of an active Janus particle partially coated with a catalytic cap promoting a chemical reaction only on a portion of its surface.

For an accurate representation of an extended active particle of finite radius, higher-order reflections in the phoretic and hydrodynamic fields should further be accounted for.
This can be achieved by including additional singularities so as to satisfy the boundary conditions imposed at the particle and at the interfaces iteratively using the classical method of successive images.
Besides, an exact solution of the phoretic and hydrodynamic problems can alternatively be obtained based on the eigensolution expansion of Laplace equation using the system of bipolar coordinates.
These aspects constitute an interesting extension of the present problem for future investigations.

\vspace{0.5cm}

\textbf{Funding.}
We acknowledge support from the Max Planck Center Twente for Complex Fluid Dynamics, the Max Planck School Matter to Life, and the MaxSynBio Consortium, which are jointly funded by the Federal Ministry of Education and Research (BMBF) of Germany and the Max Planck Society. This work was supported by Slovenian Research Agency (A.V., grant number P1-0099).

\vspace{0.25cm}

\textbf{Declaration of interests.} The authors declare no conflicts of interest.

\appendix

\section{Definition of integrals $\mathcal{L}_m^n$}
\label{appendix:Lmn}
We introduce the following integral functions of the form
\begin{equation}
	\mathcal{L}_m^n (\rho, t) = \int_0^\infty J_n(q \rho) \sin \left( q t + m \, \frac{\pi}{2} \right) \Intd q \, , 
\end{equation}
for $m, n \in \{0,1\}$.
These (improper) integrals are convergent and can be evaluated analytically.
It can be shown that for $n = 0$~\citep{abramowitz72, gradshteyn2014table},
\begin{equation} 
\mathcal{L}_0^0 (\rho, t) = \frac{\Theta (t - \rho)}{\left( t^2-\rho^2 \right)^\frac{1}{2}} \, , \qquad\qquad
\mathcal{L}_1^0 (\rho, t) = \frac{\Theta (\rho - t)}{\left( \rho^2 - t^2 \right)^\frac{1}{2}} \, , \label{K10_K20}
\end{equation}
with $\Theta (\cdot)$ denoting Heaviside's step function.
Using the fact that $J_0'(x) = -J_1(x)$ (prime denotes a derivative with respect to the argument) together with the differentiation and integration properties of trigonometric functions, it can further be shown that
\begin{equation}
\mathcal{L}_0^1 (\rho, t) = \frac{t \, \mathcal{L}_1^0 (\rho, t)}{\rho} \, , \qquad\qquad
\mathcal{L}_1^1 (\rho, t) = \frac{1 - t \, \mathcal{L}_0^0 (\rho, t)}{\rho}   \, . \label{K11_K21}
\end{equation}

\section{Evaluation of the indefinite integral in Eq.~\eqref{EqtoBeProven}}
\label{sec:appendix-integral}

In this Appendix, we show using the residue theorem in complex analysis that 
\begin{equation}
\operatorname{Im} \left\{ \int_0^\infty f(z) \, \Intd z \right\} = \frac{\pi}{2} \left( a^2+(b+c)^2 \right)^{-\frac{1}{2}} \, , \label{EqtoBeProvenAppendix}
\end{equation}
where
\begin{equation}
f(z) = z \left( z^2 + c^2 \right)^{-1} \left( a^2 + (b-iz)^2 \right)^{-\frac{1}{2}} \, , 
\end{equation}
is a complex analytical function defined in the upper half plane.
In addition, $(a,b,c) \in \mathbb{R}_+^3$.
Since $f(-z) = -\overline{f(z)}$ with bar denoting the complex conjugate and that for a given complex number~$z = x+iy$, we have $y = \left( z-\overline{z} \right)/(2i)$, it follows that
\begin{equation}
\operatorname{Im} \left\{ \int_0^\infty f(z) \, \Intd z \right\} = \frac{1}{2i} \int_{-\infty}^\infty f(z) \, \Intd z \, . \label{Eq1App}
\end{equation}

To evaluate the improper integral on the right-hand side of Eq.~\eqref{Eq1App}, we employ the usual approach by choosing a closed integration contour~$\gamma = \gamma_1 + \gamma_2$ with~$\gamma_1$ denoting the linear path along the real axis in the interval $[-R,R]$ and $\gamma_2$ the circular path of radius~$R$.
Accordingly,
\begin{equation}
\oint_{\gamma} f(z) \, \Intd z = 2i\pi \sum \operatorname{Res} (f, z_i) \, .
\end{equation}

By setting $z = R e^{i\theta}$ with $\theta = [0,\pi]$, it follows that $\Intd z = iR e^{i\theta} \, \Intd \theta$.
Then, the integral along circular path~$\gamma_2$ can be evaluated asymptotically as
\begin{equation}
\int_{\gamma_2} f(z) \, \Intd z = 2R^{-1} + \mathcal{O} \left( R^{-2} \right) \xrightarrow[R\to\infty]{} 0\, .
\end{equation}

Thus, the contour integration reduces to that along~$\gamma_1$, leading when taking the limit $R\to\infty$ to
\begin{equation}
\int_{-\infty}^\infty f(z) \, \Intd z = 2i\pi \operatorname{Res} (f, ic) = i\pi \left( a^2 + (b+c)^2 \right)^{-\frac{1}{2}} \, . \label{Eq2App}
\end{equation}

Eq.~\eqref{EqtoBeProvenAppendix} results by combining Eqs.~\eqref{Eq1App} and~\eqref{Eq2App}.

\bibliographystyle{jfm}
\bibliography{biblio,bibliodfg} 

\end{document}